\documentstyle[preprint,aps]{revtex}

\begin{document}
\title{Fluctuation, time-correlation function and geometric Phase}
\author{A. K. Pati}
\address{Theoretical Physics Division, 5th Floor, Central Complex}
\address{Bhabha Atomic Research Centre, Bombay-400 085, INDIA}
\date{\today}
\maketitle
\begin{abstract}
We  establish a fluctuation-correlation theorem by relating   the
quantum fluctuations in the generator of the parameter change
to the time integral of the quantum correlation
function between the projection operator and  force  operator  of
the  ``fast'' system. By taking a cue from linear response theory
we relate the quantum fluctuation in the generator to the  generalised
susceptibility. Relation between the open-path geometric phase,
diagonal elements of the quantum metric tensor and the force-force
correlation function is provided and the  classical limit of the
fluctuation-correlation theorem is also discussed.
\end{abstract}
~~~~~~~~~~~~~\\

~~~~~~~~~~~~~~\\
\newpage

\par
Fluctuations  in  a generic observable is inherent to all quantum
systems, which can not be  brought  to  zero  even  in  principle
unless the system is prepared in an eigenstate of the observable.
In a many body system the nature and the interrelation of quantum
fluctuation   and   statistical  fluctuation  (which  comes  from
time-correlation functions) is a topic of great interest. Given a
composite system the quantum fluctuation in  a  observable  of  a
subsystem  may  drive  the other subsystem towards equlibrium and
this information can be obtained by studying the  time-corelation
function  of  the later subsystem. One can describe the effect of
coupling between ``slow'' and ``fast''  system  by  studying  the
time-correlation  function  between different operators pertinent
to subsystems. Here, we use the concept  of  adiabatic  geometric
phase to bring out an important connection  between  the  quantum
fluctuation  and  time-correlation function of some observable of
the ``fast'' system.

\par
In adiabatic theorem, the geometric phase was discovered by Berry
\cite{1}  as  an  extra  phase shift acquired by the wavefunction
during cyclic variations of external  parameters.  Realising  its
importance,  this concept was further generalised to non-adiabatic
situations by Aharonov and Anandan \cite{2},  to  non-cyclic  and
non-unitary  situations  by  Samuel  and  Bhandari  \cite{3}  and
recently  to  non-cyclic,  non-unitary  and   non-Schr{\"o}dinger
situations  by  the  present author \cite{4}. Although, geometric
phase can appear in a  quite  general  context  \cite{5},  purely
related  to  the  geometry of the Hilbert space \cite{6}, most of
the application of the geometric phase theory deals with  systems
undergoing  adiabatic  evolutions.  Most  important  and  natural
context,  where  adiabaticity  holds  is  the  system  comprising
collection  of  electrons (fast system) and nuclei (slow system).
Here, one applies Born-Oppenheimer(BO) approximation \cite{7}  to
solve  the  slow  motion  by  integrating out the fast degrees of
system. Incidentally,  the  gauge  potential  (now  called  Berry
potential)  was  first highlighted by Mead \cite{8} as an leading
order correction to ususal  BO  approximation  prior  to  Berry's
observation.  Recently,  it  has  been  shown by Berry and Robbins
\cite{9} that there are higher order corrections to the usual  BO
force  called geometric magnetism and deterministic friction in a
classical setting and half-classical setting.

\par
In  this  letter,  we  show  that  the quantum fluctuation in the
generator of the unitary operator (which  induces  the  parameter
change)  is  directly related to the time integral of the quantum
correlation function between  the  projection  operator  and  the
force  operator  of  the  ``fast''  system. Further, invoking the
ideas of linear response theory one can show  that  this  quantum
fluctuation  is  related  to the generalised susceptibility. This
fluctuation can be represented interms of the  diagonal  elements
of  the  quantum  metric  tensor, which in turn is related to the
force-force correlation function. We provide a new expression for
the adiabatic geometric phase when the slow coordinates undergo a
non-cyclic change.  Also,  we  discuss the classical limit of the
fluctuation-correlation theorem when the classical counterpart of
the fast motion  is  both  chaotic  and  integrable.  The  formal
developments   presented   in  this  paper  will  have  important
applications in the areas such as nuclear physics  and  condensed
matters  physics  and  open up new avenues for studying classical
limit of  generalised quantum one-form for chaotic systems.

\par
Let us consider a composite, many-body system (``slow'' + ``fast'') and  denote  the
``slow'' and ``fast'' variables by $({\bf R},{\bf P})$  and
$({\bf  r},{\bf  p})$, respectively. The Hamiltonian of the total
system can be written as
\begin{equation}
H({\bf r},{\bf p},{\bf R},{\bf P}) = {{\bf P}^2 \over 2M} + {{\bf p}^2 \over 2m} +
+ V({\bf r},{\bf R}) = h({\bf r},{\bf p},{\bf R}) + {{\bf P}^2 \over 2M}.
\end{equation}
Ususally, one first solves for the fast Hamiltonian $h({\bf r},{\bf p},{\bf R})$ for  a  fixed
coordinate  of  the  slow  variable  ${\bf  R}$.  The  eigenvalue
equation reads as
\begin{equation}
h({\bf R})|n({\bf R})\rangle = \epsilon_n({\bf R})|n({\bf R})\rangle,
\end{equation}
where $|n({\bf R})\rangle$ and $\epsilon_n({\bf R})$ are the eigenstate and eigenvalue of  the  fast  system  that
depends  parametrically  on the slow variable ${\bf R}$. The wave
function of the composite system is
\begin{equation}
\Psi({\bf r},{\bf R}) = \sum_n \phi_n({\bf R}) \psi_n({\bf r},{\bf R}),
\end{equation}
with $\psi_n({\bf r},{\bf R}) = \langle{\bf r}|n({\bf R})\rangle$ and $\phi_n({\bf R})$
is the slow eigenfunction. When we integrate out the fast degrees
of  freedom,  the  effective  Hamiltonian  for  the  slow  system
contains a gauge potential ${\bf A}_n({\bf R})$, whose flux gives
the  geometric phase. Thus, the effective Hamiltonian is given by
\cite{7,8}
\begin{equation}
H_{eff}  =  {1 \over 2M}({\bf  P}  -  \hbar{\bf  A}_n({\bf  R})^2) +
\epsilon_n(
{\bf R})
\end{equation}
where ${\bf A}_n({\bf R}) = i\langle n({\bf R})|\nabla n({\bf R})\rangle$ is the
Berry  potential.  The  presence  of  gauge  potential  leads  to
observable  effects  like  shift  in  quantum  numbers  and  level
splittings  \cite{10}. If we consider the  time evolution of fast
system then during the cyclic change of the slow coordinate, the
wavefunction of the fast system acquires a geometric phase  given
by
\begin{equation}
\gamma_n(C) = i \oint_C \langle n({\bf R})|\nabla n({\bf R}) \rangle.d{\bf R}
            = \oint_C {\bf A}_n({\bf R}).d{\bf R}
\end{equation}

\par
However,  there  could  be  situations where the slow coordinates
need not undergo a cyclic variation.
It has been shown by the author \cite{11} that  when  the
parameters are adiabatically changed along
an   arbitrary  curve $\Gamma$, then the geometric phase is given
by  the  line  integral  of   a   generalised   gauge   potential
$\Omega_n({\bf R})$ as
\begin{equation}
\gamma_n(\Gamma) = i \int_{\Gamma} \langle \chi_n({\bf R})|\nabla \chi_n({\bf R})\rangle.d{\bf R} = \int_{\Gamma} \Omega_n({\bf R}).d{\bf R}.
\end{equation}
where   $|\chi_n({\bf   R})\rangle$   is   a   ``reference-eigenstate''
introduced   by   the  author,  defined  from  the  instantaneous
eigenstate as $|\chi_n({\bf R})\rangle =  {\langle n({\bf  R})|n({\bf R}(0)) \rangle
\over |\langle n({\bf R})|n({\bf R}(0))\rangle|}|n({\bf R})\rangle$. The generalised gauge potential is related to the Berry potential
as
\begin{equation}
\Omega_n({\bf R}) = {\bf A}_n({\bf R}) - {\bf P}_n({\bf R}),
\end{equation}
where ${\bf P}_n({\bf R})$ is a new gauge potential, given by
\begin{eqnarray}
& {\bf  P}_n({\bf  R}) =  {i \over 2|\langle n({\bf R}(0))|n({\bf R})\rangle|^2 } \nonumber\\
& \bigg[\langle n({\bf R}(0))|(\nabla n({\bf R})\rangle \langle n({\bf R})| - |n({\bf R}) \rangle \langle \nabla n({\bf R})|)|n({\bf R}(0))\rangle \bigg]
\end{eqnarray}

The open-path adiabatic geometric phase is  gauge  invaraint  even  if  the
nucelar coordinates do not come back to their original value.
Because   under  a  $U(1)$  gauge  transformation  of  the  fast
eigenfunction $|n({\bf  R})\rangle$ is
changed to $e^{i\alpha({\bf R})} |n({\bf R})\rangle$ wheras
${\bf A}_n({\bf R})$ and ${\bf P}_n({\bf R})$  transform as

\begin{eqnarray}
 & {\bf A}_n({\bf R})  \rightarrow & {\bf A}_n({\bf R}) - \nabla \alpha({\bf R}) \nonumber\\
 & {\bf P}_n({\bf R})  \rightarrow & {\bf P}_n({\bf R}) - \nabla \alpha({\bf R}),
\end{eqnarray}
and therefore the whole expression is gauge compensated.

\par
Let  us  focus  our  attention on the generalised gauge potential
$\Omega_n({\bf R})$. In what follows, we express it in  terms  of
the quantum fluctuation in some observable of the slow system. On
defining  a  Hermitian  operator  ${\bf  B}$ through the relation
$|\nabla n({\bf R})\rangle
= i {\bf B}|n({\bf R})\rangle$, we can express the potential ${\bf P}_n({\bf R})$ as

\begin{equation}
{\bf   P}_n({\bf   R})  =  -{1  \over  2}\bigg[  {\langle  n({\bf
R}(0))|{\bf  B}|n({\bf   R})   \rangle   \over   \langle   n({\bf
R}(0))|n({\bf  R})  \rangle} + {\langle n({\bf R})|{\bf B}|n({\bf
R}(0))\rangle \over \langle  n({\bf  R})|n({\bf  R}(0))  \rangle}
\bigg]
\end{equation}
Using the fact \cite{12} that the action of any Hermitian
operator  $O$ on some state $|\Psi\rangle$ can be written as $O|\Psi\rangle ~ =
\langle O \rangle |\Psi \rangle + \Delta O|\Psi_{\bot}\rangle$, where $\langle O \rangle$ is the average
and $\Delta O = \sqrt(< O^2> - <O>^2)$ is the  uncertainty  in  the
operator $O$, respectively. The state $|\Psi_{\bot} \rangle$
belongs  to  the  orthogonal   complement subspace of the Hilbert
space, such that  $\langle \Psi|\Psi_{\bot}\rangle  =  0$.  For  the  adiabatic
eigenstate and the operator ${\bf B}$, we have
\begin{equation}
{\bf  B}|n({\bf  R}) \rangle =  -{\bf A}_n({\bf R})|n({\bf R})\rangle + \Delta
{\bf B}|n_{\bot}({\bf R})\rangle.
\end{equation}
where $\langle n({\bf  R})|{\bf  B}|n({\bf  R}) \rangle =  -{\bf A}_n$.
With the help of the above equation, the potential
${\bf P}_n({\bf R})$ can be expressed as
\begin{equation}
{\bf P}_n({\bf R}) = {\bf A}_n({\bf R}) -
 Re \bigg({\langle n({\bf R}(0))|n_{\bot}({\bf   R})\rangle   \over  \langle n({\bf
R}(0))|n({\bf R})\rangle } \bigg) \Delta {\bf B}
\end{equation}

This shows that the Berry potential  is  just  a  part  of  the  gauge
potential   ${\bf    P}_n({\bf   R})$.   Hence,  the  generalised
potential  can  be  expressed  in terms of the fluctuation in the
operator ${\bf B}$ as
\begin{equation}
\Omega_n({\bf R}) =  Re \bigg({\langle n({\bf R}(0))|n_{\bot}({\bf   R}) \rangle   \over  \langle n({\bf
R}(0))|n({\bf R})\rangle } \bigg) \Delta {\bf B}
\end{equation}
This  shows  that  the  open-path geometric phase acquired by the
fast eigenstate is related  to  the
integral of the fluctuation in the operator ${\bf B}$.

On the other hand the generalised gauge  potential  can  also  be
expressed as a time-correlation function. To arrive at this, we
relat it to the matrix elements of product of
projection operators and force operator (force operaor is $- \nabla
h({\bf   R})$).   Let  us  first  introduce  a  complete  set  of
eigenstates in the expression for ${\bf P}_n({\bf
R})$. Then we obtain
\begin{eqnarray}
& {\bf  P}_n({\bf  R}) =
-Im\langle n({\bf  R})|\nabla n({\bf  R}) \rangle  - \nonumber\\
& Im
\sum_{m  \not=  n}  { \langle n({\bf  R}(0))|m({\bf R}) \rangle \over
\langle n({\bf R}(0))|n({\bf  R})\rangle }  { \langle m({\bf  R})|\nabla
h|n({\bf R})\rangle \over
             (\epsilon_n({\bf R}) - \epsilon_m({\bf R}))}
\end{eqnarray}
where we have used the equality $\langle n|\nabla m \rangle =  {\langle n|\nabla  h|m \rangle
\over (\epsilon_n - \epsilon_m)}$ for $m \not= n$. Since ${\bf A}_n =
-Im\langle n({\bf    R})|\nabla   n({\bf    R})\rangle$,   we  can  write  the
generalised vector potential as
\begin{eqnarray}
& \Omega_n({\bf  R}) =
{1 \over |\langle n({\bf R}(0))|n({\bf  R})\rangle|^2} \nonumber\\
& Im
\sum_{m  \not=   n}   {\langle n({\bf  R})|P_n({\bf  R}(0))P_m({\bf  R})
\nabla h |n({\bf  R})\rangle \over
(\epsilon_n({\bf R}) - \epsilon_m({\bf R}))}
\end{eqnarray}
where  $P_n({\bf   R}(0))$  and $P_m({\bf  R})$ are instantaneous
projection operators corresponding  to  the  eigenstate  $|n({\bf
R}(0))\rangle$  and  $|m({\bf   R})\rangle$,  respectively.  On comparing two
expressions (13) and (15) for the gauge potential we obtain

\begin{eqnarray}
& Im \sum_{m  \not=   n}   {\langle n({\bf  R})|P_n({\bf  R}(0))P_m({\bf  R})
\nabla h |n({\bf  R})\rangle \over
(\epsilon_n({\bf R}) - \epsilon_m({\bf R}))} = \nonumber\\
& Re \bigg(\langle n({\bf R}(0))|n_{\bot}({\bf R}) \rangle \langle
n({\bf R})|n({\bf R}(0)) \rangle \bigg) \Delta {\bf B}.
\end{eqnarray}
Further, we simplify the left hand side of the  above  expression
using an integral representation of the energy denominator, viz
${1  \over  (\epsilon_n  -  \epsilon_m)} = {1 \over \hbar} lim_{s
\rightarrow 0} \int_0^{\infty}~dt e^{-st} \sin((\epsilon_n  -  \epsilon_m){t \over \hbar}) $.

Therefore, we have
\begin{eqnarray}
& {1 \over \hbar} lim_{s \rightarrow 0} \int_0^{\infty} ~dt ~e^{-st}  \nonumber\\
& Im \sum_{m  \not=   n} \sin((\epsilon_n  -  \epsilon_m){t \over \hbar})  \langle n({\bf  R})|P_n({\bf  R}(0))P_m({\bf  R}) \nabla h |n({\bf  R}) \rangle
= \lambda({\bf R}) \Delta {\bf B}.
\end{eqnarray}

where $\lambda({\bf R}) = Re( \langle n({\bf R}(0))|n_{\bot}({\bf R}) \rangle \langle n({\bf
R})|n({\bf R}(0)) \rangle )$ is a real sacle factor.
Then  define  a  quantum   correlation   function   between   the
instantaneous projection operator and the force operator as
\begin{equation}
Q(t) = {1 \over 2}\langle n|(A_{-t}B + BA_{-t}) - (AB_t + B_tA)|n
\rangle
\end{equation}
where $A = P_n(0)$ and $B = \nabla h$  and $A_{-t}$ is time-evolved
operator of $A$ (with $t$ replaced by  $-t$)  and  $B_t = (\nabla
h)_t$  is  the
time-evolved operator of $\nabla h$ at fixed ${\bf R}$. A similar
quantum correlation function and its various moments have been
studied  in different context \cite{13}  while  discussing  the  quantum-classical
discordance for chaotic systems. It can be shown that the quantum
correlation function defined above is precisely what we  have  in
left hand side, i.e.,
\begin{eqnarray}
& Q(t) = -2 Im
\sum_{m  \not=   n}  sin ((\epsilon_n  -  \epsilon_m){t \over \hbar}) \nonumber\\
& \langle n({\bf R})|P_n({\bf  R}(0))P_m({\bf  R})
\nabla h |n({\bf  R})\rangle.
\end{eqnarray}
which is an almost periodic function,  only  if  $\{\epsilon_n\}$
forms  a  complex sequence whereupon the convergence of such sums
have to be treated nicely \cite{apf}.
Thus, we arrive at our first result
\begin{equation}
-{1 \over  2\hbar} \int_0^{\infty} dt~Q(t)  = \lambda \Delta {\bf
B}
\end{equation}
where   we   have   left  the  convergence  factor  implicit.

To provide a physical meaning for the fluctuation in operator ${\bf
B}$  we  consider a family of Hamiltonians $h({\bf R})$ which are
unitarily   related   to   a   unparametrised   Hamiltonian   $H$
\cite{rob}, i.e.

\begin{equation}
h({\bf R}) = U({\bf R}) H U({\bf R})^{\dagger}
\end{equation}

Then one can define the  generators of the unitary operator $U({\bf
R})$ as ${\bf g}({\bf R}) = i\hbar \nabla U({\bf R}) U({\bf R})^{\dagger}$.
These unitary operators $U({\bf R})$ not necessarily constitute a
group  and  provide  a connection between the parameter dependent
eigenbasis $|n({\bf R})>$ and parameter-independent  basis  $|n>$
as  defined  through  $|n({\bf  R})>  =  U({\bf R})|n>$. Thus the
generator of the parameter ${\bf g}$ is nothing but the operator
$-\hbar {\bf B}$. Therefore, the quantity  $\Delta  {\bf  B}$  is
related  to  the  fluctuation  in  the generator of the parameter
depenedent unitary operator.

It  may  be  worth  recalling  that  the  usual  Berry  potential
represents the average of the generator whereas  the  generalised
gauge potential represents the fluctuation in the said generator.
Thus (20) {\it relates the time integral of a quantum correlation
function  of  the  fast  system to the quantum fluctuation of the
generator of the unitary operator}- the  central  result  of  our
letter.  This  may  be  called  a fluctuation-correlation theorem
analogous to the fluctuation-dissipation theorem  in  statistical
mechanics.

\par
Another meaning can be favoured for the  above  expression.  Note
that  the  quantum  correlation  function $Q(t)$ defined above is
actually a difference  of  two  symmetrised  time-correlation
functions, i.e.,
$Q(t) = C_{AB}(-t) - C_{BA}(t)$ where
$C_{AB}(t)  = {1 \over 2}\langle n|(A_{-t}B + BA_{-t})|n \rangle$ and $C_{BA}(t) =
{1\over 2}\langle n|(AB_t + B_tA)|n \rangle$. Appealing to linear response
theory of adiabatic many-body quantum system  \cite{14}  we  can  define  a
generailsed  susceptibility  in terms of the Laplace transform of
the symmetrised time-correlation function, which is given by
\begin{equation}
\chi_{AB}(z) = \int_0^{\infty} e^{-zt} C_{AB}(t) dt
\end{equation}
With this idea (20) can be expressed, alternatively as
\begin{equation}
lim_{z \rightarrow 0} [\chi_{AB}(z) - \chi_{BA}(z)] = - 2 \lambda \Delta {\bf P},
\end{equation}
This relates the quantum fluctuation to the susceptibility of the system
in  the  limit  $z  \rightarrow  0$,  which  is of statistical in
nature.

\par
Next,  we  express  the  open-path geometric phase interms of the
quantum metric tensor.
First we show that  the
uncertainty in $B_i, (i = 1,...N)$ is nothing  but  the  diagonal
elements  of  the  positive  semi-definite  quantum metric tensor
$g_{ij}$, where $g_{ij}$ is given by

\begin{equation}
g_{ij} = Re\langle \partial_i n|\partial_j n \rangle - (i\langle n|\partial_i n \rangle) (i\langle n|\partial_j n \rangle)
\end{equation}

The  physical significance of the $g_{ij}$ is that it defines the
distance \cite{15,16} between any two points along an arbitrary  path  in  the
parameter  space corresponding to the evolution of the eigenstate
in  the  Hilbert  space.
To  see  this  recall  that  the  infinitesimal distance function
between the adiabatic eigenstate $|n({\bf R}) \rangle$ and $|n({\bf R}+ d{\bf R}) \rangle$
is given by
\begin{eqnarray}
ds^2  & = & \bigg[ \langle \partial_i n|\partial_j n \rangle - (i\langle n|\partial_i n \rangle) (i\langle n|\partial_j n \rangle) \bigg]d{\bf R}_id{\bf R}_j \nonumber\\
      & = & T_{ij}d{\bf R}_id{\bf R}_j  = g_{ij}d{\bf R}_id{\bf R}_j
\end{eqnarray}
where $T_{ij} = g_{ij} + iv_{ij}$ is the Hermitian quantum geometric tensor,
$g_{ij}$ is the  real  symmetric  tensor  and  $v_{ij}$  is  real
antisymmetric    tensor.    The   quantum   geometric  tensor  is
manifestly gauge
invariant. The antisymmetric tensor field is nothing
but  the phase two-form which gives the adiabatic Berry phase. The
real symmetric tensor gives us the distance between  the  quantum
states.  But  interestingly,  the  generalised  vector  potential
(phase one-form) is related to the diagonal elements of the  real
symmetric tensor.
The  diagonal  elements  describe   the
uncertainties because $(\Delta B_i)^2 = g_{ii}$ and off-diagonal elements describe the correlations
between the operators $B_i$'s.
With the help of these metric structures we can recast the
geometric phase as
\begin{equation}
 \gamma_n(\Gamma)   =  \int  {\lambda  \over  |<n({\bf R})|n({\bf
R}(0))>|^2}  \surd g_{ii}({\bf R}).dR_i
\end{equation}
An  immediate  interpretation  of  the  above  result is that the
open-path geometric phase for the ``fast'' system is the integral  of
the  scaled symmetric tensor, during an arbitrary, adiabatic evolution of a
``slow'' quantal system.

\par
Further,  the  diagonal elements of the real symmetric tensor can
be expressed as a force-force correlation function which is a very
useful quantity in studying its classical  limit  when  the  fast
motion is chaotic \cite{13,19}. On writing $g_{ii}$ as
\begin{equation}
g_{ii} = Re \sum_{m  \not=  n}  {\langle n| \partial_i h |m \rangle
\langle m| \partial_i h |n \rangle
\over
(\epsilon_n  - \epsilon_m)^2}
\end{equation}
and using the integral representation of the energy term, viz
$(\epsilon_n    -   \epsilon_m)^{-2}   =   -{1   \over   \hbar^2}
\int_0^{\infty}  dt~t  exp(i{(\epsilon_n   -  \epsilon_m)t  \over
\hbar})$ we have
\begin{equation}
g_{ii} = -{1 \over 2\hbar^2} \int_0^{\infty} dt~t [ \langle n|(\partial_i h)_t
(\partial_i  h)|n  \rangle  + \langle n|(\partial_i h)(\partial_i
h)_t|n \rangle
\end{equation}
where  $(\partial_i h)_t$ is the time-evolved operator. Since the
geometric  phase  is  related  to  the metric structure it is also
related  to  force-force   correlation   function,   where   same
components  of  the force operators are involved. However, we can
not say that the geometric phase is related  to  friction in pure
quantum  mechanics.  Because,  friction  in  such  situations  is
related  to  the  symmetric  part  of the force-force correlation
function (where different components of the force  operators  are
involved).  One  can  see  in  \cite{18} that for chaotic quantum
systems the phase two-form is related to the anti-symmetric  part
of   the  force-force  correlation  function.  Since  for  cyclic
evolution of the  adiabatic  parameters  our  result  \cite  {11}
reduces  to the result of Berry \cite{1}, it is natural to expect
that  the  one-form  is  in  some  way  related  to   force-force
correlation function.

\par
Before  concluding  this  paper,  we  briefly discuss some issues
related to the classical limits of this results.  One  can  study
the classical limit of the fluctuation-correlation relation, when
the fast motion  is  chaotic.  This  is  important,  because  the
quantum fluctuations and correlations of large systems which have
classical chaotic manifestations is of great interest.  When  the
classical  counterpart  of  the  fast system is chaotic we assume
that the mixing property holds. In  this
case  the quantum expectation values of physical quantities
correspond to the phase space  average  over  the  microcanonical
distribution on invariant energy surface, \cite{17,18,19} i.e.,
\begin{equation}
\langle  n  |O| n \rangle  \rightarrow  <O>_E  =  {  \int d^N{\bf r}
d^N{\bf p}~ O({\bf  r},{\bf  p},{\bf  R})  ~\delta(E  -
h({\bf R})) \over \int d^N{\bf r} d^N{\bf p} ~\delta(E -
h({\bf R})) }
\end{equation}
The  time-evolved  operators  in  quantum  case correspond to the
physical quantities at time  evolved  points  on  the  trajectory
generated  by  the  fast Hamiltonian $h({\bf R})$. Therefore, the
classical  analog  of  fluctuatio-correlation  theorem   can   be
expressed as
\begin{equation}
\int_0^{\infty} dt~Q_c(t) = - 2 \hbar \lambda_c <({\bf B}  -<{\bf
B}>_E)^2>_E
\end{equation}
where
$Q_c(t) = {\int d^N{\bf r} d^N{\bf p}
~\delta(E - h({\bf R})) (A_{-t}B -B_t A) \over
\int d^N{\bf r} d^N{\bf p}
~\delta(E -  h({\bf  R}))  }$,  is the classical valued
correlation function and  $\lambda_c$  is  the  classical  valued
scale  factor.
Recently, it has been shown by Srednicki \cite{20} that the  time
variation of the quantum fluctuation in the any observable can
be   interpreted  as  appropriate  thermal  fluctuation  in  that
observable when the number of degrees of freedom  $N$  is  large.
Therefore,   the   fluctuation  in the generator of the parameter
change can  be regarded as a
thermal fluctuation in classical statistical sense.
The  classical and quantum correlation functions
can be useful in studying the various moments \cite{13} and  in  analysing
the  clash  of  limits  $\hbar \rightarrow 0$ and $ t \rightarrow
\infty$ which may  shed  some  light  on  the  behaviour  of  the
geometric  phase  for  classical  chaotic  systems. However, such
studies are beyond the scope of this letter. One can remark  that
when  the  classical  counterpart  of the fast system is chaotic,
mixing property would imply that at  long  time  ($t  \rightarrow
\infty$) the classical correlation function vanishes. This in turn
imples  that  the fluctuation property of the slow system as seen
by  the  fast  system  also  vanishes.  Therefore,  to  see   the
statistical  fluctuation  property  of  the slow system, study of
long time behaviour of the chaotic trajectory is not prefered.

\par
If the classical counterpart of the fast  system  is  integrable,
then  the quantum expectation values can be replaced by the torus
average of the  classical  valued  function  over  the surface  of
constant action ${\bf I}$, i.e.,
\begin{equation}
\langle  n |O| n \rangle  \rightarrow  <O>_I  =  {1 \over (2\pi)^N} \int d^N \theta
 ~~O({\bf  r},{\bf  p},{\bf  R})  ~\delta({\bf I}  -
{\bf I}({\bf r},{\bf p},{\bf R}))
\end{equation}
and the fluctuation correlation theorem takes the form
\begin{equation}
\int_0^{\infty} dt~Q_c(t) = - 2\hbar \lambda_c <({\bf  B}  -<{\bf
B}>_E)^2>_I
\end{equation}
where
$Q_c(t) = {1 \over (2\pi)^N} \int d^N \theta
~\delta({\bf I} - {\bf I}({\bf r},{\bf p},{\bf R})) (A_{-t}B -B_{t} A)$.

Although   these  discussions  are  pertained  to  collection  of
electrons (fast) and nuclei (slow), they are equally valid in the
situations where one can separate the motion or energy  sacle  to
two  regimes  and  BO approximation holds. Recently, we \cite{21}
have  attempted  to  understand   the   damping   of   collective
excitations   in   finite-Fermi  systems  using  the  concept  of
adiabatic geometric phase and response functions, when the system
is fully chaotic.

\par
To conclude this paper,  we  have studied the adiabatic geometric
phase  acquired  by  the  ``fast''  system  when   the   ``slow''
coordinates   undergo   a   non-cyclic   variation,   within  the
Born-Openheimer setting. This resulted in the  formulation  of  a
fluctuation-correlation    theorem,    which    says   that   the
time-integral of the correlation function of the ``fast''  system
is  proportional  to  the quantum fluctuation of the generator of
the parameter change of the ``slow'' system as  measured  in  the
``fast''  system.  The fluctuation in the corresponding generator
can be related to the generalised susceptibility of the  ``fast''
system. Invoking the idea of geometric distance function, we have
related  the  quantum  one-form  to  the diagonal elements of the
quantum  metric  tensor.  Also,  the  classical  limit   of   the
fluctuation-correlation theorem is discussed when the fast motion
is  chaotic  and integrable. This work opens up the possiblity of
studying the  spectral  one-form  $F(\epsilon)$  defined  through
$F(\epsilon)  =  \sum_n  \delta (\epsilon - \epsilon_n) \Omega_n$
and  in  answering  the  classical  limit  of  quantum   one-form
$\Omega_n$ for chaotic systems in future.\\

~~~~~~~~~~~\\

{\bf Acknowledgements}:  I  would  like  to  thank the referee for
suggesting  me  to correct the interpretation for the fluctuation
in the oepartor ${\bf B}$.


\begin{thebibliography}{99}
\bibitem{1} M. V. Berry, Proc. R. Soc. London, {\bf A 392} (1984) 457.
\bibitem{2} Y. Aharonov and J. Anandan, Phys. Rev. Lett. {\bf 58} (1987) 1593.
\bibitem{3} J. Samuel and R. Bhandari,  Phys. Rev. Lett. {\bf 60} (1988) 2339.
\bibitem{4} A. K. Pati, J. Phys. A, {\bf 28} (1995) 2087.
\bibitem{5} N. Mukunda and R. Simon, Ann. Phys. {\bf 228}  (1993)
20.
\bibitem{6} A. K. Pati, Phys. Lett. A {\bf 202} (1995) 40.
\bibitem{7}  R.  Jackiew,  Int.  J.  Mod. Phys. A {\bf A} {\bf 3}
(1988) 285.
\bibitem{8} C. A. Mead, J. Chem. Phys. {\bf 70} (1979) 2284.
\bibitem{9}  M. V. Berry and J. M. Robbins, Proc. R. Soc. London,
{\bf A 442} (1993) 659.
\bibitem{10} A. shapere and  F.  Wilczek  (Eds.)  {\sl  Geometric
phases in Physics}, (World Scientific, Singapore, 1989).
\bibitem{11}  A.  K.  Pati, {\sl Adiabatic Berry phase and Hannay
angle for open paths}, (preprint, 1996).
\bibitem{12}  Y.  Aharonov  and L. Vaidman, Phys. Rev. A {\bf 41}
(1990) 11.
\bibitem{13} J. M. Robbins and M. V. Berry, J. Phys. A {\bf 25} (1992) L961.
\bibitem{apf} B. Jessen and H.  Tornhave,  Acta  Math.  {\bf  77}
(1945) 138.
\bibitem{rob} J. M. Robbins, J. Phys. A {\bf 27} (1994) 1179.
\bibitem{14} R. Kubo, M.  Toda  and  N.  Hashitsume,  Statistical
Physics-II (Springer Verlag, Tokyo, 1978).
\bibitem{15} J. P. Provost and G.  Vallee,  Commun.  Math.  Phys.
{\bf 76} (1980) 289;
\bibitem{16} A. K. Pati, Phys. Lett. A {\bf 159} (1991) 105.
\bibitem{17}  Verdiere  Y  de Colin, Compositio Mathematica, {\bf
27} (1973) 83; 159.
\bibitem{18} M. V. Berry and J. M. Robbins, Proc. R. Soc. London,
{\bf 442} (1993) 641.
\bibitem{19}  J. M. Robbins and M. V. Berry, Proc. R. Soc. London,
{\bf 436} (1992) 631.
\bibitem{20} M. Srednicki, J. Phys. A {\bf 29} (1996) L75.
\bibitem{21} S. R. Jain and A. K. Pati, Phys. Rev. Lett. {\bf 80}
(1998) 650.

\end{thebibliography}
\end{document}